*Experimenta coram societate instituta in confirmationem theoriae pressionum quas latera canalis ab aqua tranfluente sustinet*

Translated and Annotated
by
Sylvio R. Bistafa[*]
February 2019

Foreword

In the 1729 issue of *Commentari Academiae Scientiarum Imperialis Petropolitanae* (Vol. IV, 1729 (1735), pp. 194-201), Daniel Bernoulli (1700-1782) publishes: *Experimenta coram societate instituta in confirmationem theoriae pressionum quas latera canalis ab aqua tranfluente sustinet* ('Experiments made before the instituted society in confirmation of the theory of the pressures which the walls of the channel sustain due to the flowing water'). In this publication, Daniel describes six experiments conducted before the St. Petersburg Academy, in confirmation of his theory, which had been published two years earlier in the 1727 issue of *Commentari Academiae Scientiarum Imperialis Petropolitanae* (Vol. II, 1727 (1729), pp. 111-125): *Theoria nova de motu aquarum per canales quoscunque fluentium* ('New theory on the water motion flowing through whatever channels'). This publication is dedicated to the motion of water through ducts, and contains the developments that led to the relation between the pressure of the flowing water and the amplitude of the duct, which translates into what became later known as Bernoulli's principle, which states that an increase in the speed of a fluid occurs simultaneously with a decrease in pressure. Both works were included by Daniel in his famous book *Hydrodynamica*[1], which was published in 1738. In this book, the five experiments were extended to include four more experiments (in a total of 10 experiments), and also includes the ideas that appeared in *Theoria nova*.

---

[*] Corresponding address: sbistafa@usp.br
[1] Bernoulli, D. (1968). Hydrodynamics. Jointly translated with Bernoulli, J. Hydraulics from the Latin by T. Carmody and H. Kobus. Dover Pub. Co. See also the translated and annotated version by Ian Bruce at http://www.17centurymaths.com/contents/dbernoullicontents.htm

As Calero[2] points out, the concept of pressure remained as a property linked to hydrostatics, independent of motion, until Daniel Bernoulli demonstrated experimentally that this supposition was not true, and that the pressure on the walls of a discharge tube depends on the velocity of the water in the duct. As we shall see, to this end, Daniel used the piezometer, which consists of a vertical tube through which the water ascends according to the pressure in its base. When the flow in the duct is blocked, the level of the water in the tube reaches the height of the fluid in the reservoir. However, when the flow in the duct is reestablished, the pressure will be lowered, which is evident from the drop in the tube level, and it will go down as the flow in the duct increases. The experiment has several remarkable points: it shows the relation between pressure and velocity, introduces the piezometer as an instrument to measure pressure − still widely used until today −, and presents experimental results which will promote the theoretical development required to explain it.

Although the experiments described by Daniel in *Experimenta coram societate* appear in the two translated versions of *Hydrodynamica*, the objective of the present work is to give a first known translation of the original manuscript in Latin, which contains the first experimental confirmation of one of the most important basic principles in fluid mechanics.

'Experiments made before the instituted society in confirmation of the theory of the pressures which the walls of the channel sustain due to the flowing water'

by

Daniel Bernoulli   Son of Johann

What are the pressures of the waters in a filled vessel is known by everyone since remote times. However, as far as I know, so far nobody has correctly defined the pressure of the flowing waters through channels: there has been those who have considered the argument that the walls of the channels on which the waters are stagnant or flowing are, in both cases, equally pressed in the same way, with a magnitude corresponding to the altitude of the waters: but this is false. During the interim time, a very fitting argument was found by me, which should be examined with great attention. Thence, it will be considered the estimation of the real quantity of water, which are conveyed through tubes implanted on the laterals of channels, which, according to the work of *Frontinus,* and after him many others: it is also uncertain the necessary strength of the aqueducts, and many other questions related to the solution of the motion of fluid through the animal body, and likewise other such questions. However, soon I realized that our problem cannot be solved until the motion of the waters will had been correctly defined, and thus, in fact, should be defined at every moment, since from rest, or at the beginning of the motion, until a given maximum acceleration had been known: what is the reason why so far this has been concealed, how the fluids through channels drive the motion, pressure; not only that I have uncovered, except previously in the general theory *on the motion of the waters through channels flowing whithersoever*, which we see in *Comment. Acad. Scient. Petrop. Tom. II pag. 111.* With the power of that theory I provided the complete Statics of the motion of fluids, of which I will appoint here certain examples as well as experimental results publically confirmed by the society: nor this follows the strait paths of the sort prescribed in the Academic Dissertations, and for this reason I choose to allow another approach, to collect in a single treatise, all that I have meditated upon the subject of waters.

---

[2] Calero, J. S. (2008). The Genesis of Fluid Mechanics 1640-1780. Series: Studies in History and Philosophy of Science, Vol. 22, p.24. Springer.

Imagine, therefore, a cylindrical tube connected horizontally to a reservoir of infinite amplitude and full of water, in which an external orifice is blocked by my finger: thus we see that according to the ordinary rules, each part of the tube is subjected to a pressure given by the total height of the water. However, with the finger removed, the waters begin to flow through the tube, and soon the pressure of the water will be reduced, until eventually it will fade away: and yet it will not change instantaneously, particularly in the longer tubes: until finally it will fade away, with all the water acquiring the possible velocity; it is only after an infinite time, if we wish to speak precisely, in fact as quickly as to accelerate the motion of the water, so that I say that the velocity can rise as much as possible fast enough not all at once, unless the waters in the reservoir fills entirely a very long duct.

If indeed the waters through the tube do not rush out with the orifice entirely filled, the pressure of the waters will not be raised up completely, although the waters flow with the full velocity through the tube.

We will not pay attention to those pressures changes, which take place at the beginning of the flow, until the flow of the waters is suppose to reach uniformity: we say at least with the final pressure which the flow reaches uniformity: also impediments which can retard the flow of the waters through the tube, such as adhesions of the water in the laterals of the tube, contractions of the outflow observed by *Newton* (which is equally fortuitous and cannot be all together avoided) and others similar ones for which we do not have an explanation.

Be now the amplitudes of the tube and its orifice discharging waters such as $m$ to $n$, the height of the water in the reservoir above the tube $= a$: be understood that the water occupies the whole orifice and has its full velocity, which is dictated by the height $a$, flowing; also the pressure of the water with the orifice obstructed as if proportional to the height of the water be indicated by $a$: I say that the pressure of the water flowing through the tube will be equal to $\frac{mm-nn}{mm}a$, and, therefore, zero, if $m = n$, that is, in case the tube is completely open.

For the experimental confirmation of the Static of fluid motions, I used a wooden box, whose width was equal to one foot, the length equal to three feet, the height equal to fourteen inches: I filled this [box] with water, and I implanted horizontally, in its lower part, a cylindrical tube accurately fabricated from iron: and in fact that tube was made of iron.

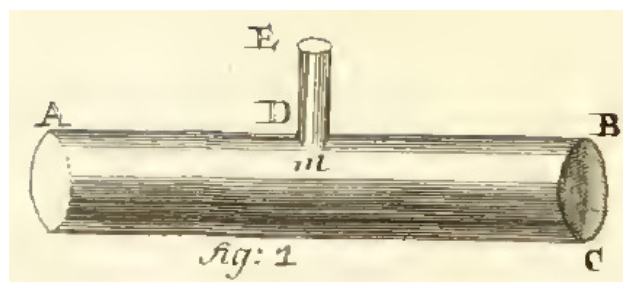

Certainly it had a length *AB* of 4 *English* poll., 2 lin., a Diameter *BC* of 7. lin.; in the middle of the tube a small hole $m$ was perforated, and in that place a small tube *DE* likewise of iron and having six lin. in length, and having one and a half lin. in diameter has been welded, and with the small hole $m$ at the middle of its base.

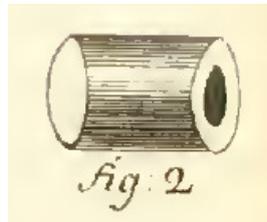

fig: 2

From this small tube, afterwards, I placed a tube made of glass of equal amplitude, just as it appears in the third figure, which shows the whole experimental apparatus. Furthermore, I arranged to produce three covers adapted to the iron tube, and perforated with an opening of different magnitude: such a cover is depicted in the third figure.

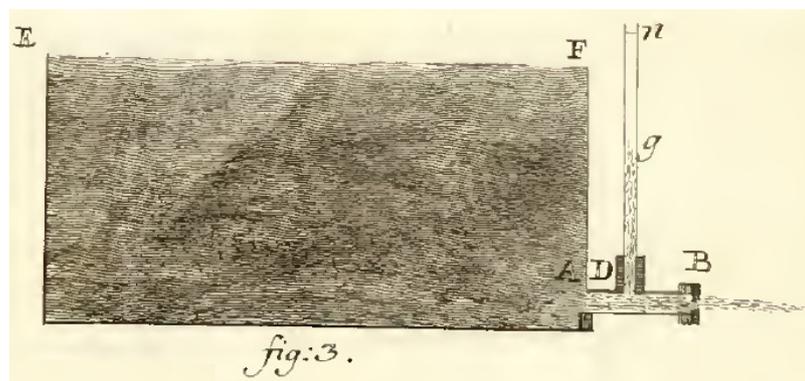

fig: 3.

With all these combined in the manner which the third figure shows, and provided that no water flows through other cracks but only through the aperture $BC$, I obstructed the orifice in $BC$, and at the same time I observed point $n$ in the glass tube laid in the vertical, to which the waters have ascended, which I marked rolling around a silken string: but first, I had investigated the capillary effect on this glass tube, and this I had found to be five lines[3], such that the vertical water tube allowed a difference between each surface of the water consisting of five lines: therefore, the point $n$ was elevated above the surface *EF* just as many lines, and hence in the calculation of any height $Dn$, $Dg$ five lines is agreed to be deducted. In each experiment, the box was thus maintained filled with water, so that the height $AF$ was 9 poll., 7 lin., whereas the height $Dn$ was 10 poll.[4] Thus, with all these prepared for the experiment, and with the orifice at $BC$ having been opened, the efflux of the water was allowed, and immediately the water descended in the glass tube, such as from $n$ to $g$, which location $g$ we marked again with another silken string circumvolving the tube as before. And thus, finally, we carried on the experiments.

---

[3] i.e. $\frac{5th}{12}$ of an inch.

Experiment 1. When the diameter of the aperture in the cover $BC$ was $2\frac{1}{5}$ lin. the descent $ng$ was a little more than one line, so that no difference between the theory and the outcome of the experiment will be possible to be observed.

Experiment 2. With another cover taken, in which the diameter of the aperture was $3\frac{2}{5}$ lin., or a little greater, the observed descent $ng$ was six lines and two thirds, clearly again as the theory indicates.

Experiment 3. With the third cover adopted, in which the diameter of the aperture was five lines, or somewhat less: we observed a descent $ng$ of 28 lines. As per force of the theory it should be around 29 lines, even though the aperture was seen not have the entire five lines in diameter. The small difference is attributed to hindrances, which the water suffers in flowing through the tube, in greater amount than in the preceding experiments, in face of the enlarged motion within the tube.

Experiment 4. Finally, with no cover, we allowed the waters to flow freely through the opening, and then nearly all the water in the glass tube was discharged: yet, some part remained, which we observed to be eight lines high. However, of these, five should be attributed to the efect of the capillary tube: the remaining three are attributed to hindrances, which the water strikes during flow from $D$ to $B$.

---

[4] 1 *pollicare* (poll.) = 12 *lineas* (lin.). With no flow, $AF$ should be equal to $Dn$. $AF$ was maintained at 9 poll. and 7 lin. $Dg$ was 10 poll., which by subtracting 5 lin. due to the meniscus height formed at the free surface of the water in the glass tube (capillary effect), gives 9 poll. and 7 lin., and hence, equal to $AF$, as it should be. Finally, one *pollicare* (1 poll.) would correspond to one inch (1 in).

Thus, the experiments done exactly agree with the theory. Thence, moreover, it is not difficult to see that it can happen that the laterals of the tube are not only pressed towards the exterior, but also that they are compressed inwards towards the axis of the tube[5]: but this was deduced from another experiment.

Experiment 5. I placed a conical cylindrical tube $AB$, whose outlet orifice was greater than the inlet orifice, and simultaneously I employed a curved glass tube, which is shown in figure 4.

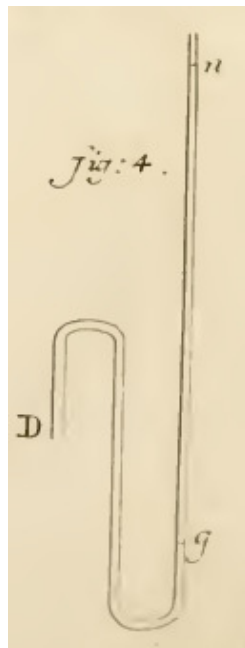

And since before flow, the water in the glass tube clung to $n$, it descended in this tube all the way to $g$, and since the water was flowing through a conical tube, it happened that point $g$ was below $D$; proving that it was being compressed during flow through the conical tube. Thus, also for these cases, the hindrances to the motion are remarkable, which make the water velocities in the outlet orifice much less than those corresponding to the height of the water: and this because the height of point $D$ above $g$ was not as great as otherwise it would be: nonetheless there was some [height].

A similar effect I obtained in another way, but even more remarkable; this other experiment I performed in the following year in the presence of the Academics; the Most Serene

---
[5] This refers to pressure changing into suction inside the tube.

Prince *Emanuel* of Portugal being present [According to Ian Bruce the Price had visited the St. Petersburg Academy in 1730: *Id. Ibid.*, p. 1].

Experiment 6. In figure 5, $ACFB$ represents a cylinder, in the bottom of which, was implanted the conical tube $DGHE$; and this had a small tube at the side at $l$, which received the extremity of a curved glass tube $lmn$, the height $CA$ was 3 poll., 10 lin., E$l$ 4 lin, $l$H 2 poll., $9\frac{1}{2}$ lin., the amplitude of the conical tube in $l$ was to the amplitude of the orifice $GH$ as 10 is to 16, the height $ln$ was 5 poll., 6 lin. and its orifice $n$ was submerged in the water in the small vessel $M$.

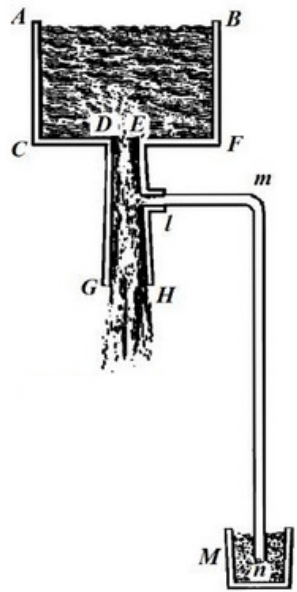

Fig. 5

Putting the finger over the orifice $GH$ and the vessel filled, the water dripped through the glass tube $lmn$ into vessel $M$: but once the finger was removed and with the water now flowing through $GH$, the water, in a reciprocal motion, climbed up spontaneously from the small vessel $M$ through the tube $nml$, and uniting with the remaining water, flowed out through $GH$, until the entire vessel $M$ had been emptied. However, the water was continuously supplied above, so that the vessel was kept full. If part of the orifice was obstructed by a finger, it was easy to make the waters in the glass tube $lmn$ to be moved up or down at will. It was remarkable to see in that experiment that the placed vessel $M$ was much smaller than the orifice $GH$.

Thence, it appears that the reason why the smoke that rises through the smoke stack not only does not exit through the openings, which are usually made in the sides of the smoke stack, but also draws the air itself behind with great motion: It should be also noted that the sides of the tube $DGHE$ to be compressed towards the interior, even if this tube were cylindrical and the smoke stack were upside down, since the smoke elevates itself because of its fluidic nature.

Moreover, in matters that pertain to the static of the fluid motion, there are some topics that still require another theory: Also in any case where the motion of the waters must be correctly defined previously, which is safely determined according to the same pressure: which is demonstrated by another example, we shall consider a vertical cylinder of almost infinite length, having in the middle a horizontal diaphragm, and also in its lower part an equally horizontal diaphragm, and then, these diaphragms will be perforated with equal or unequal holes. If this cylinder is full of water, and the lower orifice is closed, it is clear that each part of the cylinder is pressed according to the ordinary rules: However, as soon the water begins to flow through the holes of even very small amplitudes, the amplitude of the cylinder comes into play, and other will be the pressures in different parts of the cylinder, which are determined by the lower diaphragm: and often the parts nearest to the diaphragm will be pressed inwards; in fact those [parts] near the upper diaphragm, conserve their pressures.

Thus, also, if in $A$ (Fig. 3) another cover is installed, another will be the pressure at the walls of the tube $AC$, and also, another will be the velocity of the water outflow in $BC$. This may be seen in any arrangement, however they are not within the scope of the above theory, that with its assistance, it is easy to define the pressure and the velocity: however, I encouraged very much experiments of any kind, which always confirmed the strength of the conceived theory: That were not in fact [confirmed] before the instituted society, here their reassessment will not linger.